\documentclass{elsart5p}
\pdfoutput=1
\usepackage[utf8]{inputenc}              
\usepackage[T1]{fontenc}                  
\usepackage{xcolor}                       
\usepackage{amsfonts}
\usepackage{amsmath}
\usepackage{amssymb}
\usepackage[english]{babel}
\usepackage{graphicx}
\usepackage{units}
\usepackage{url}                          
\usepackage[hang,raggedright]{subfigure}  
\usepackage{nicefrac}
\usepackage{textcomp}                     
\usepackage[pdfpagelabels]{hyperref}
\newcommand{\dd}{\textrm{d}}
\newcommand{\ee}{\textrm{e}}
\newcommand{\cawo}{CaWO$_{4}$ }

\DeclareMathAlphabet{\mathbi}{OML}{cmm}{b}{it}

\allowdisplaybreaks[1]

\definecolor{grey}{rgb}{0.75,0.75,0.75}
\definecolor{brown}{rgb}{0.5,0.25,0.0}
\definecolor{pink}{rgb}{1.0,0.5,0.5}
\definecolor{darkgreen}{rgb}{0,0.5,0}
\definecolor{orange}{rgb}{1,0.5,0}
\definecolor{plotorange}{rgb}{1,0.65,0}
\definecolor{lightgray}{rgb}{0.8,0.8,0.8}
\definecolor{gold}{rgb}{1,0.8,0}
\definecolor{superfluid}{rgb}{0.5,0.75,0.75}
\definecolor{normalliquid}{rgb}{0,0.5,0.5}

\hypersetup{
  pdftitle={Glued CaWO4-Detectors for the CRESST-II Experiment},
  pdfauthor={Michael Kiefer},
  pdfkeywords={Dark Matter, WIMP, Solid-state detectors, Low temperature detectors, Scintillation detectors, CaWO4, Epoxy, Glue}
}

\begin{document}

  \begin{frontmatter}
    \title{Glued \cawo Detectors for the CRESST-II Experiment}
    \author{Michael Kiefer\corauthref{Kiefer}\ead{kiefer@mppmu.mpg.de}}
    \author{, Franz Pröbst, Godehard Angloher, Irina Bavykina, Dieter Hauff, Wolfgang Seidel}
    \address{Max-Planck-Institut für Physik, München, Germany}
    \corauth[Kiefer]{Corresponding author. Address: Max-Planck-Institut für Physik, Föhringer Ring 6, D-80805 M\"unchen, Germany. Tel: +49 +89 32354 237, Fax: +49 +89 32354 526}
    \received{6 June 2008}
    \revised{15 September 2008}
    \accepted{?? September 2008}
    \received{6 June 2008}
\revised{15 September 2008}
\accepted{29 September 2008 }
\begin{abstract}
      The Cryogenic Rare Event Search with Superconducting Thermometers Phase II (CRESST-II) at the L.N.G.S in Italy is searching for Dark Matter using low-temperature calorimeters. These detectors allow to discriminate different particles by simultaneous measurement of phonons and scintillation light. The sensors used consist of superconducting tungsten thin-film thermometers, which measure the thermal effect of the phonons created in an attached absorber crystal. It has been observed that the scintillation  of the \cawo absorber degrades during the process of depositing the tungsten film. In order to prevent this, a new technique for producing the detectors was investigated. This technique might also be valuable by expanding the range of scintillator materials suitable for producing a Dark Matter detector.
    \end{abstract}
  \end{frontmatter}

  \section{Dark Matter Search with CRESST-II}
    \subsection{Introduction}
      Since the observations of Zwicky in the 1930's the nature of Dark Matter remains unknown. Experiments like WMAP~\cite{Spe06} could only deliver more evidence on its existence but could not clarify its nature. The weakly interacting massive particle (WIMP) is a well-motivated candidate for Dark Matter~\cite{jungman}.\par
    \subsection{CRESST-II}
      The CRESST-II experiment aims for direct detection of WIMPs  scattering off nuclei in a scintillating absorber crystal~\cite{Ang05}. The target material is chosen in order to optimize the cross section for spin-independent coherent WIMP-nucleus scattering, whose cross section scales quadratically with the atomic mass $A$.\par
      The energy transfered by a scattering is typically only in the order of \unit[10]{keV}. This, together with an extremely low interaction rate of less than 10 events per kilogram of absorber material and year of exposure requires very sensitive detectors, which allow the rejection of background events~\cite{jungman}.\par
\begin{figure}
        \centering
        \includegraphics[width=6cm]{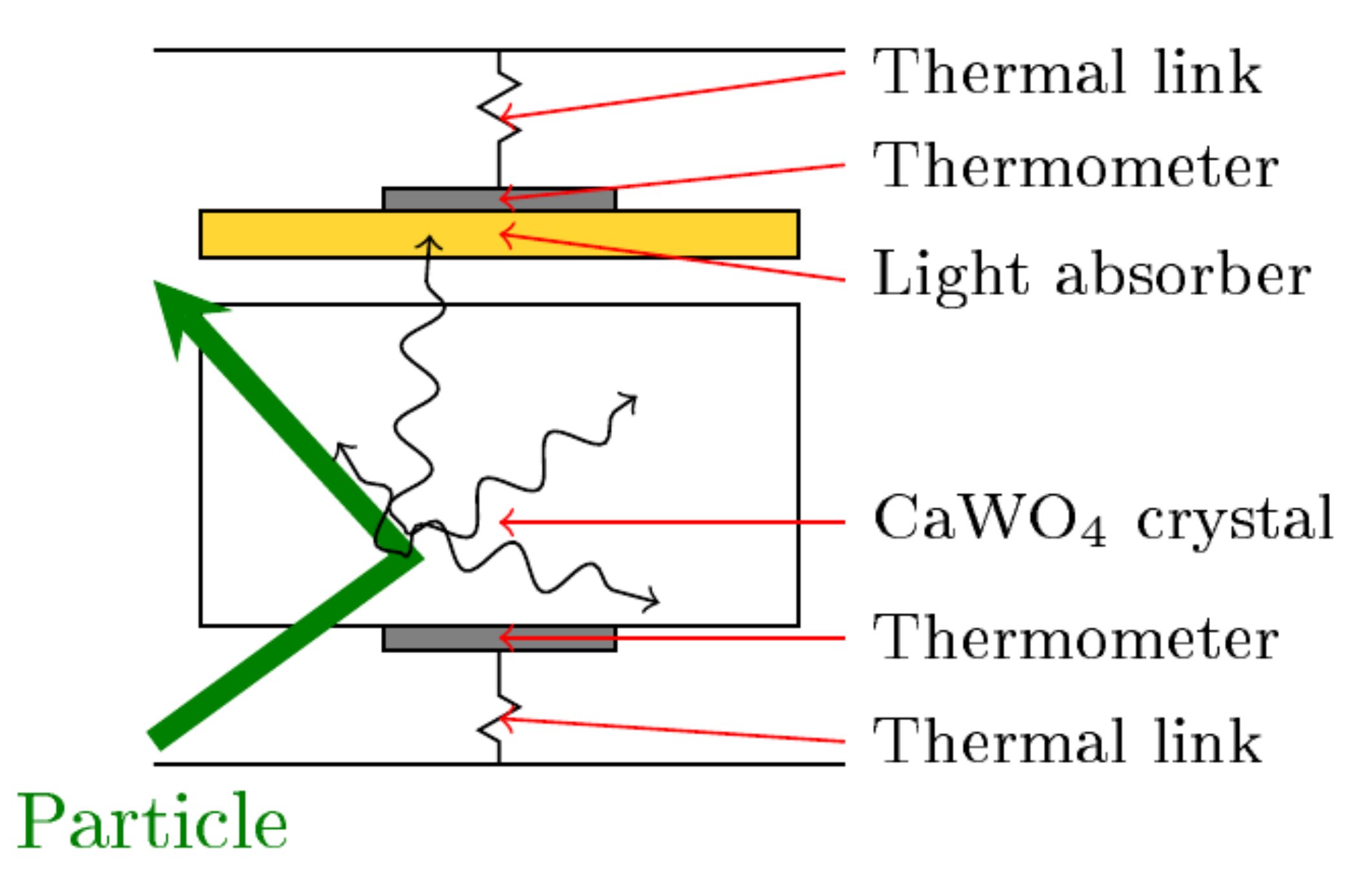}
        \caption{Schematic view of the detector.}\label{fig:schemaDetector}
      \end{figure}

      In order to reject the background radiation, CRESST-II uses scintillating crystals as target material. A schematic view of the detector can be seen in Figure \ref{fig:schemaDetector}. The energy deposited in a scintillating absorber crystal by a particle interaction excites both phonons and scintillation light. The relative amount of light depends on the nature of the detected particle. As WIMPS, in comparison to highly ionizing particles, cause a relatively low light output, the optimization of detection of the scintillation light is of utmost importance.\par
      The scintillation light is absorbed by a silicon coated sapphire crystal, exciting phonons in the crystal lattice. Each of the two crystals of a detector module, \cawo and sapphire, is connected to a superconducting phase transition thermometer for measurement of the phonon signals.\par
      Such a thermometer consists of a thin film of superconducting material (tungsten) which is deposited on the crystal. Phonons are absorbed in the film and increase the temperature of its electron system. The film is thermally stabilized in its phase transition from the superconducting to normal conducting state, therefore its resistance is extremely sensitive to variations in temperature. Phonons that are absorbed in the film increase the temperature of its electron system and thus its resistance. The change in resistance is read out by a sensitive SQUID circuit~\cite{Ang05,Antonio,Lang:2008fa}.\par
      The upcoming EURECA experiment is going to use particle detection and discrimination techniques developed and studied in the CRESST-II experiment~\cite{EureKraus}.
  \section{Glued Detectors}
    The idea behind the use of glued detectors is to produce the superconducting phase transition thermometers on small substrates that later on are glued onto the scintillating absorber crystals. This has several advantages:
    \begin{description}
     \item[Light yield:] For the evaporation of a tungsten film, the scintillating absorber crystal is heated up. Heat, however, degrades the light output of the scintillator~\cite{Ninkovic05}, resulting in a decreased particle discrimination capability.
     \item[New materials:] Other attractive scintillator materials may suffer even more from the W-deposition and etching. Producing the thermometer on a small substrate avoids the exposition of most of the scintillator material to these treatments. In this way, the gluing technique already allowed the use of a ZnWO$_{4}$ scintillating absorber crystal in the current run~\cite{IrinaCryoScint}.
     \item[Mass production:] Producing several thermometers at once on a single substrate increases the overall speed of detector production. The substrate can be cut and the thermometers then can be glued onto several absorber crystals. For the upcoming \mbox{EURECA} experiment which aims at increasing the target mass, many more detector modules are needed than the 33 foreseen for CRESST-II.
     \end{description}
    \subsection{Proof-of-principle experiment}
\begin{figure}
        \centering
        \includegraphics[width=8cm]{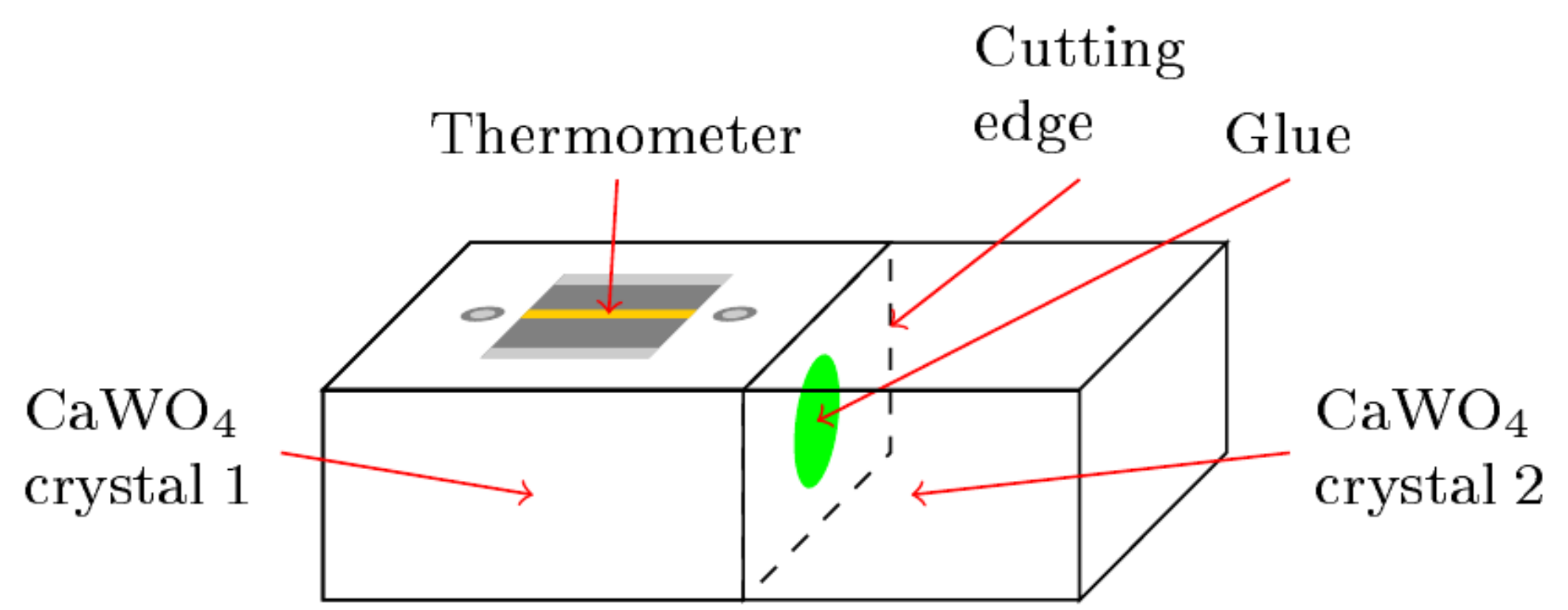}
        \caption{Schematic view of the proof-of-principle experimental setup.}\label{fig:C14setup}
      \end{figure}

      For the investigation of glued detectors, a proof-of-principle experiment was performed. This experiment was not shielded from background radiation. In order to cope with the relatively high count rates, a setup smaller than the one used at Gran Sasso was chosen: Instead of a cylindrical crystal of $\unit[40]{mm}$ height and $\unit[40]{mm}$ diameter, a cuboid of $\unit[20\times10\times5]{mm^{3}}$, named C14 was used (see Fig. \ref{fig:C14setup}).\par 
      The experiment consisted of two parts: First, the whole cuboid \cawo crystal carrying a superconducting phase transition thermometer was exposed to $\unit[60]{keV}$-gamma radiation from an $^{241}$Am source.\par
      Then, the crystal was cut into two halves, one of them carrying the thermometer.These two halves were glued together with Araldite 2011\textregistered\footnote{Araldite is a registered trademark of Huntsman Corporation or an affiliate thereof}, a two-component epoxy resin. Signals from a \unit[60]{keV} gamma source were measured, once with the radiation collimated to the half without thermometer and secondly with both halves uniformly irradiated.
    \subsection{Theoretical model}
      A theoretical model~\cite{proebst} of the behaviour of a one-crystal-setup with a superconducting phase transition thermometer was extended in order to describe the more complex case of a glued detector.\par
      The detection of phonons works by the following principle: Absorption of radiation in the scintillating absorber crystal creates high-frequency phonons that do not thermalize on the millisecond time scale. The phonons travel through the crystal ballistically before being collected in the thermometer. There, the phonon energy is transferred into the electron system of the superconducting phase transition thermometer, changing its electrical resistance by a measurable amount.
      \subsubsection{Single absorber crystal}
\begin{figure}
          \centering
          \includegraphics[width=5.5cm]{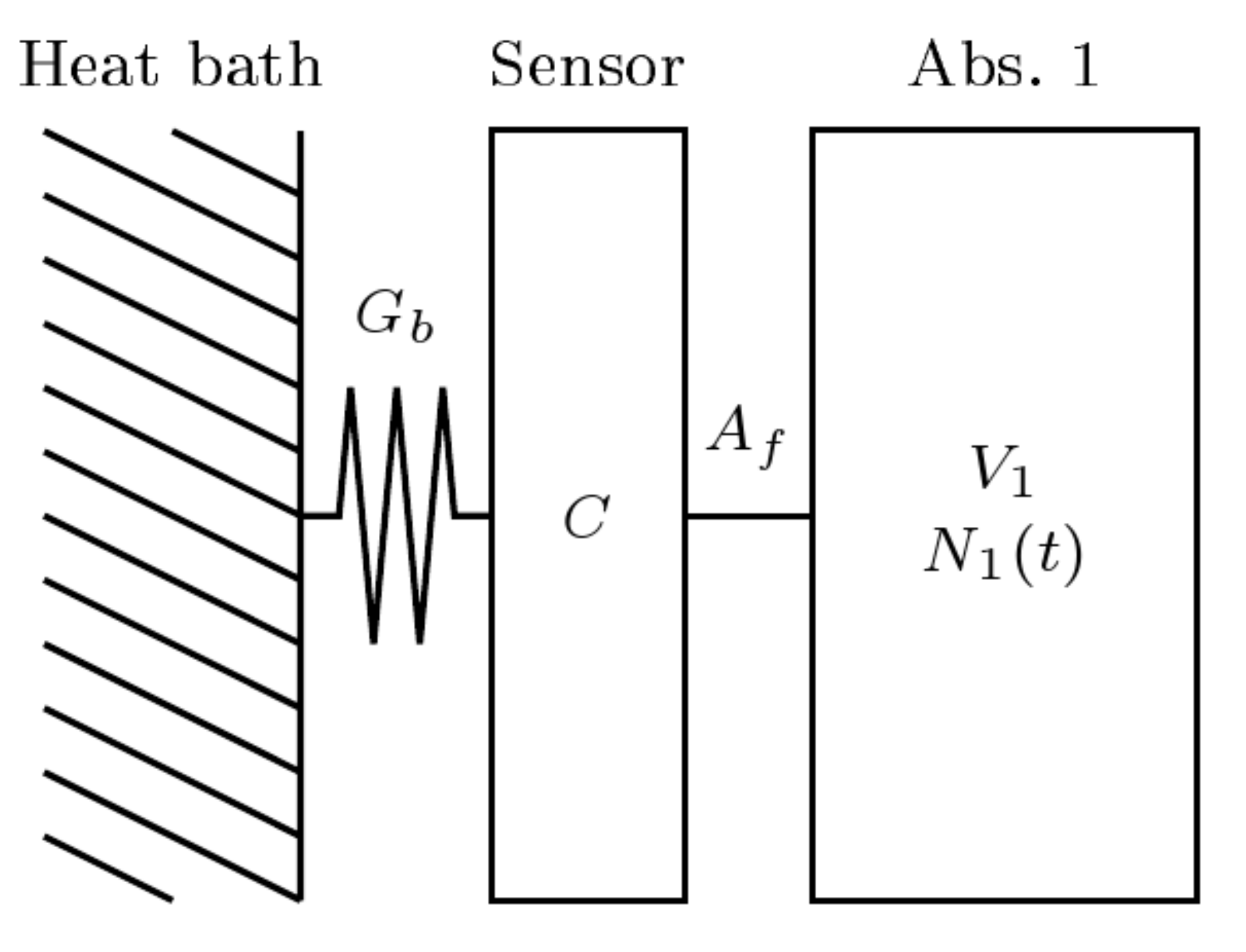}
          \caption{Schematic view of a one-crystal-setup: $V_1$, $N_1$: Volume and Phonon population in the crystal; $A_f$: Parameter for transition of phonons between crystal and thermometer; $C$, $G_b$: Heat capacity of the thermometer and coupling to the heat bath.}\label{fig:schemaOneCrystal} 
        \end{figure}

        The model makes the following assumptions: The crystal has a volume $V_{1}$ and a high-frequency phonon population $N_{1}(t)$. The transfer of non-thermal phonons into the superconducting phase transition thermometer is described by the transition parameter $A_{f}$. The superconducting phase transition thermometer itself has a heat capacity $C$ and is linked with a conductance $G_{b}$ to a heat bath of constant temperature (see Fig. \ref{fig:schemaOneCrystal}).
        Neglecting other losses, this means that the phonon population $N_1$ in the absorber crystal is determined by the deposited energy $E$, the energy per phonon $\mathcal{E}$ and the rate $\nicefrac{\dd N_1}{\dd t}$ at which the phonons flow into the thermometer.
        \begin{equation}
          \frac{\dd}{\dd t} N_{1}(t) = - A_{f} \frac{N_{1}(t)}{V_{1}}.\label{eqn:OneXtalRates}
        \end{equation}
          The phonons enter the thermometer from the absorber crystal and leave through the thermal link. This raises the temperature of the thermometer by an amount of $\Delta T$ above its equilibrium temperature:
        \begin{align}
          \frac{\dd}{\dd t} \Delta T(t) C &= -\frac{\dd}{\dd t} N_{1}(t) \mathcal{E}- G_{b} \Delta T(t)\\
          &= A_{f} \frac{N_{1}(t)}{V_{1}}\mathcal{E} - G_{b} \Delta T(t)\label{eqn:deltaT} 
        \end{align}
        By solving these two differential equations with the initial conditions of $N_{1}(t=0)=\nicefrac{E}{\mathcal{E}}$ and $\Delta T(t=0)=0$ one obtains a pulse consisting of two exponentials describing the temperature of the thermometer:
        \begin{equation}
          \Delta T(t)=\frac{A_{f}E}{A_{f}C-G_{b}V_{1}}\left(\ee^{-\frac{G_{b}}{C}t}-\ee^{-\frac{A_{f}}{V_{1}}t}\right)\label{eqn:ModelSingle}
        \end{equation}
        The quantities $\nicefrac{G_{b}}{C}$ and $\nicefrac{A_{f}}{V_{1}}$ are considered as the decay time $\tau_{n}$ and the rise time $\tau_{r}$ of the pulse, respectively.
      \subsubsection{Two absorber crystals}
\begin{figure}
          \centering
          \includegraphics[width=8cm]{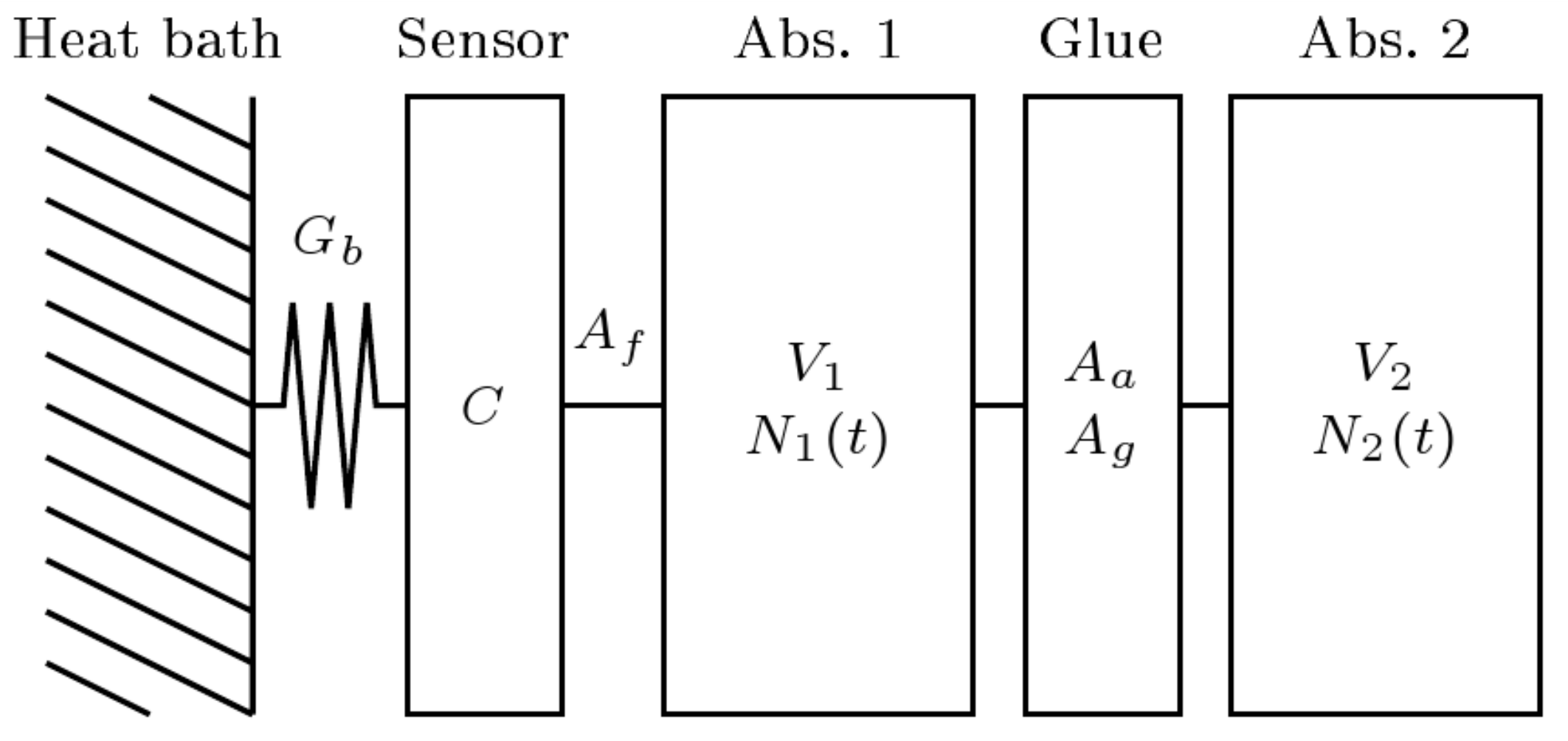}
          \caption{Schematic view of a glued-detector setup: $V_x$, $N_x$: Volume and Phonon population in the crystals; $A_f$, $A_g$, $A_a$: Parameter for transition of phonons between absorber crystal 1 and thermometer, transition and absorption through the glue; $C$, $G_b$: Heat capacity of the thermometer and coupling to the heat bath.}\label{fig:schemaTwoCrystals}
        \end{figure}

        The extension (see Fig. \ref{fig:schemaTwoCrystals}) of the model for the glued detector consists of two parts: There is a second absorber crystal which  has a volume $V_{2}$ and a phonon population $N_{2}(t)$. It is attached to the first one by a layer of glue. There are possibilities for a phonon in the glue to either pass through or to be absorbed, described by the transition parameter $A_{g}$ and the absorption parameter $A_{a}$. Phonons reflected by the glue do not have to be accounted separately, as they do not affect the balance of the two populations.\par
        Mathematically, this is represented by the following equations:
        \begin{align}
          &\frac{\dd}{\dd t} N_{1}(t) &=& - A_{f} \frac{N_{1}(t)}{V_{1}} - A_{g} \left(\frac{N_{1}(t)}{V_{1}}-\frac{N_{2}(t)}{V_{2}}\right)\notag\\
          & & &- A_{a} \frac{N_{1}(t)}{V_{1}}\label{eqn:N1}\\
          &\frac{\dd}{\dd t} N_{2}(t) &=& - A_{g} \left(\frac{N_{2}(t)}{V_{2}}-\frac{N_{1}(t)}{V_{1}}\right)- A_{a} \frac{N_{2}(t)}{V_{2}}\label{eqn:N2}
        \end{align}
        The left hand side and the first term of the right hand side of Equation \eqref{eqn:N1} are the same as in the one-crystal model (see Equation \eqref{eqn:OneXtalRates}). Additionally the phonons leaving one crystal and the phonons entering from its counterpart are taken into account by the $A_g\left(\dots\right)$ term. The last term treats phonons absorbed by the glue. The second crystal does not carry a thermometer, therefore Equation \eqref{eqn:N2} lacks the term coupling with $A_{f}$.\par
        The temperature rise of the thermometer film in the two-crystal case is again described by Equation \eqref{eqn:deltaT}.
        It makes sense to solve this system of three coupled differential equations for two different cases of initial conditions:
        If an event is caused by a particle in the first absorber then $N_{1}(t=0)=\nicefrac{E}{\mathcal{E}}$ and $N_{2}(t=0)=0$. Alternatively, if the particle is absorbed in the second crystal then $N_{1}(t=0)=0$ and $N_{2}(t=0)=\nicefrac{E}{\mathcal{E}}$. For both cases, $\Delta T(t=0)=0$, as in the one-crystal-experiment.\par
        The resulting temperature signals are of the form
        \begin{align}
           &\Delta T(t)=\notag\\
           &\alpha_{n}\left[\ee^{-\frac{G_{b}}{C}t}-\left(\frac{\gamma_{n}}{\phi}\sinh \left(\phi t\right)+\cosh \left(\phi t\right)\right)\ee^{-\beta t}\right]\label{eqn:ModelAComplex}
        \end{align}
        which still has the characteristics of a two-exponential-pulse. The different initial conditions result in different values for the $\alpha_{n}$ and $\gamma_{n}$. They are, as well as the parameters $\gamma$, $\beta$ and $\phi$, functions of the different couplings and volumes:
        \begin{align}
          \phi=&\frac{\sqrt{\left(\left[A_{f}+A_{a}+A_{g}\right]V_{2}-\left[A_{a}+A_{g}\right]V_{1}\right)^{2}+4A_{g}^{2}V_{1}V_{2}}}{2V_{1}V_{2}}\\
          \beta=&\frac{\left(A_{f} + A_{a} + A_{g} \right) V_{2} + \left( A_{a} + A_{g} \right) V_{1} }{2 V_{1} V_{2}}\\
          \begin{split}
            \gamma_{1}=&\frac{\left(A_{f}+A_{a}+A_{g}\right)V_{2}-\left(A_{a}+A_{g}\right)V_{1}}{2V_{1}V_{2}}\\
            &+\frac{A_{g}^{2}C}{\left(G_{b}V_{2}-\left[A_{a}+A_{g}\right]C\right) V_{1}}
          \end{split}\\
          \begin{split}
            \alpha_{1}=&A_{f}E\left[G_{b}V_{2}-\left(A_{a}+A_{g}\right)C\right]\\
            &/ \{ \left(G_{b}^{2}V_{1}-\left[A_{f}+A_{a}+A_{g}\right]G_{b}C\right)V_{2}\\
            &+\left(A_{a}+A_{g}\right)\left(\left[A_{f}+A_{a}\right]-G_{b}CV_{1}\right)+A_{a}A_{g} \}
          \end{split}\\
          \gamma_{2}=&\beta-2\frac{G_{b}}{C}V_{1}\\
          \begin{split}
            \alpha_{2}=&A_{f}EA_{g}C\\
            &/\{ \left(G_{b}^{2}V_{1}-\left[A_{f}+A_{a}+A_{g}\right]G_{b}C\right)V_{2}\\
            &+\left(A_{a}+A_{g}\right)\left(\left[A_{f}+A_{a}\right]-G_{b}CV_{1}\right)+A_{a}A_{g} \}
          \end{split}
        \end{align}
    \subsection{Results}
      \subsubsection{Pulse shape discrimination}
\begin{figure}
          \centering
          \includegraphics[width=7cm]{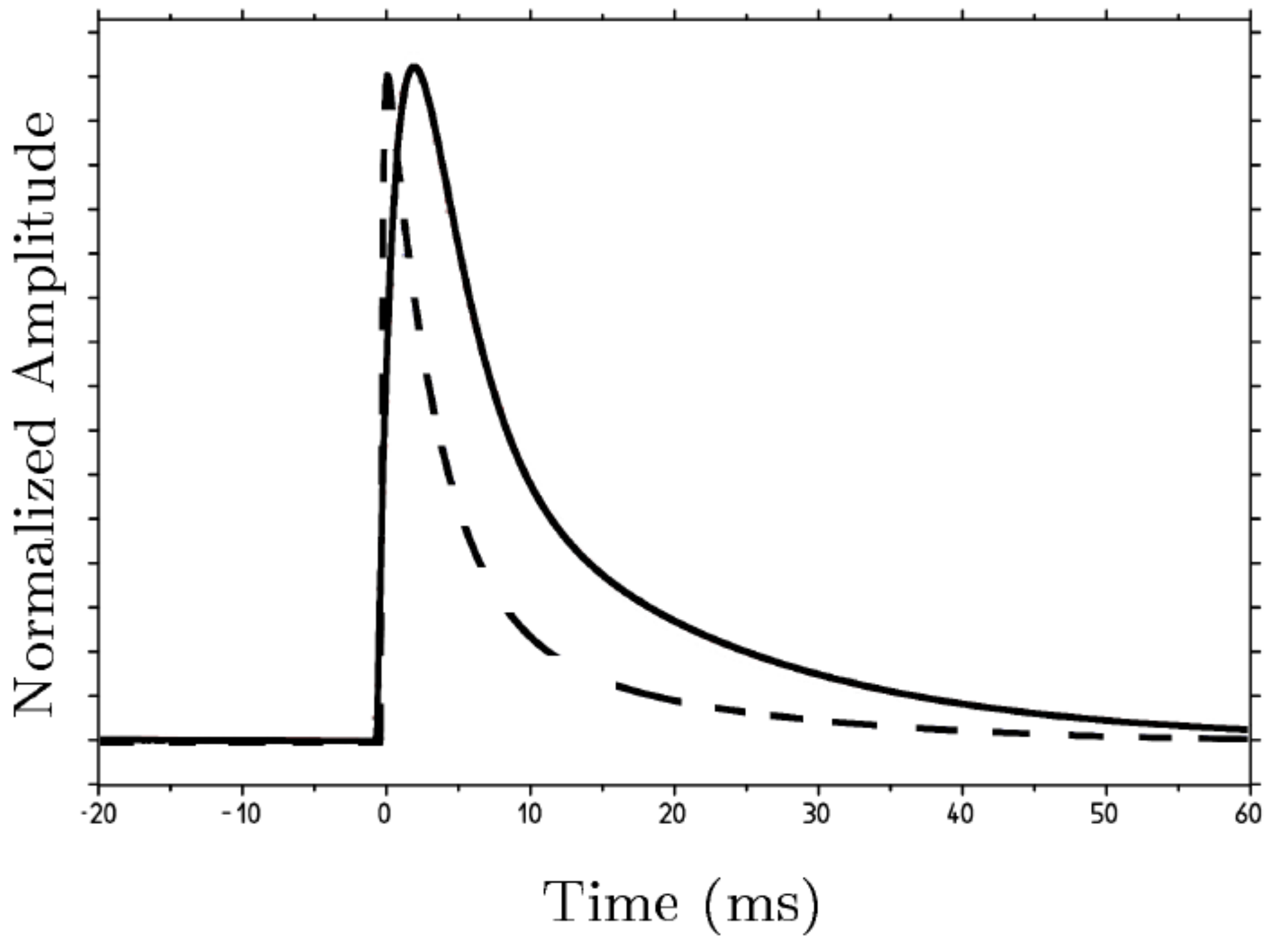}
          \caption{Fit functions for average pulses with fast rise time (dashed), resulting from an event in the thermometer carrier and with slow rise time (normal), resulting from an event in the glued part. Pulse heights are not in scale.}\label{fig:C14both}
        \end{figure}

        The temperature signal  from the thermometer was fitted using the model of the one-crystal-setup (see Equation \eqref{eqn:ModelSingle}) for the one-crystal-setup as well as for the glued detector, in order to clean the data from pile-ups. In case of the cut crystal, two distinct classes of pulses (see Fig. \ref{fig:C14both}) have been recorded which differ in rise time by a factor of 10. This is consistent with the results in~\cite{jean2}.\par
        Using the information of the collimated and uncollimated measurements, it was possible to attribute the faster pulses to events in the thermometer carrier and the slower pulses to events in the glued part without thermometer.
      \subsubsection{Spectra}
\begin{figure}
          \centering
          \includegraphics[width=8cm]{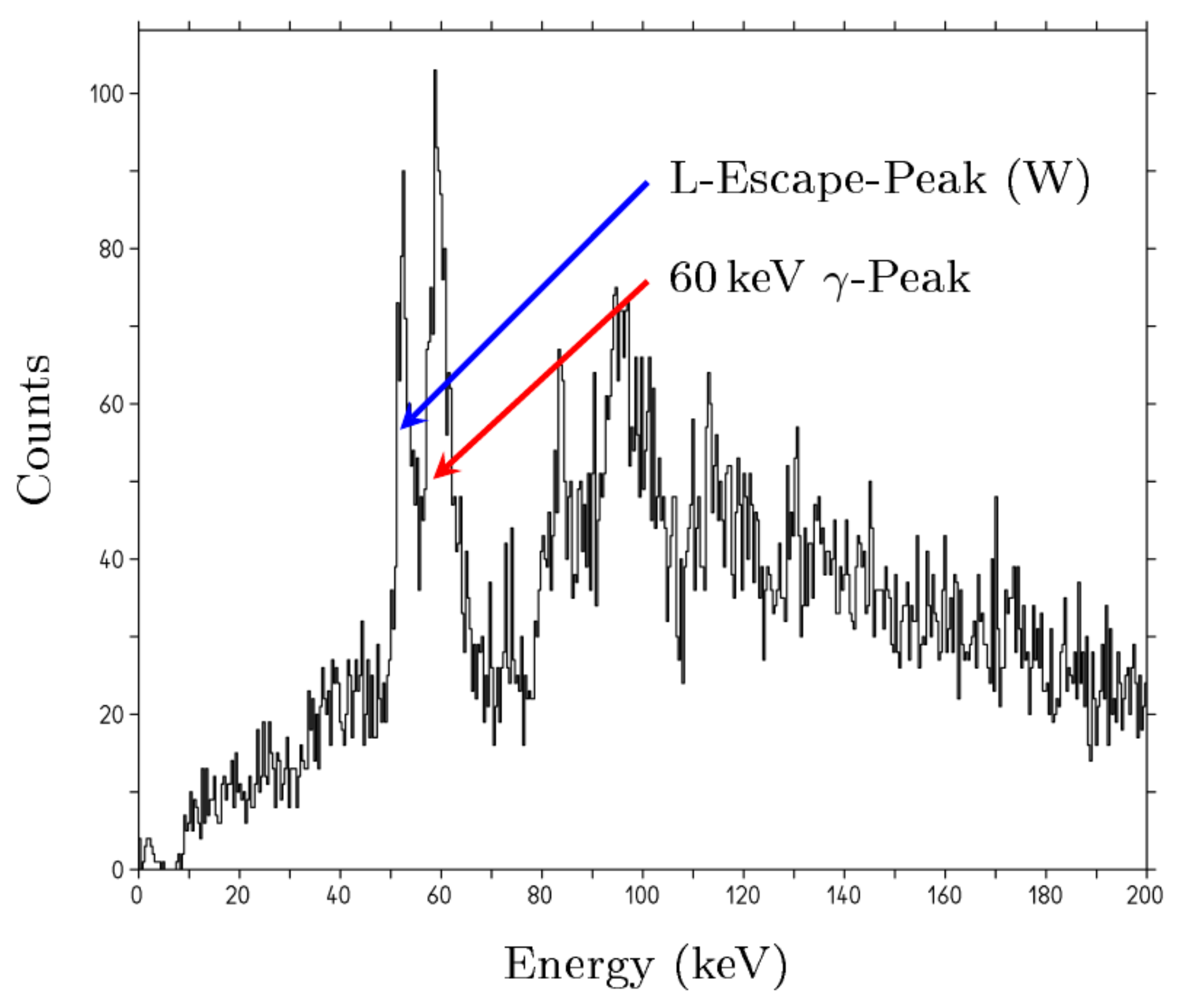}
          \caption{Spectrum taken with the uncut C14 crystal. The resolution is $\approx \unit[7.5]{\%}$ for the $\unit[60]{keV}$ peak.}\label{fig:C14uncut}
        \end{figure}

\begin{figure}
          \begin{center}
            \subfigure[Fast pulses (from thermometer carrier)]{
              \includegraphics[width=8cm]{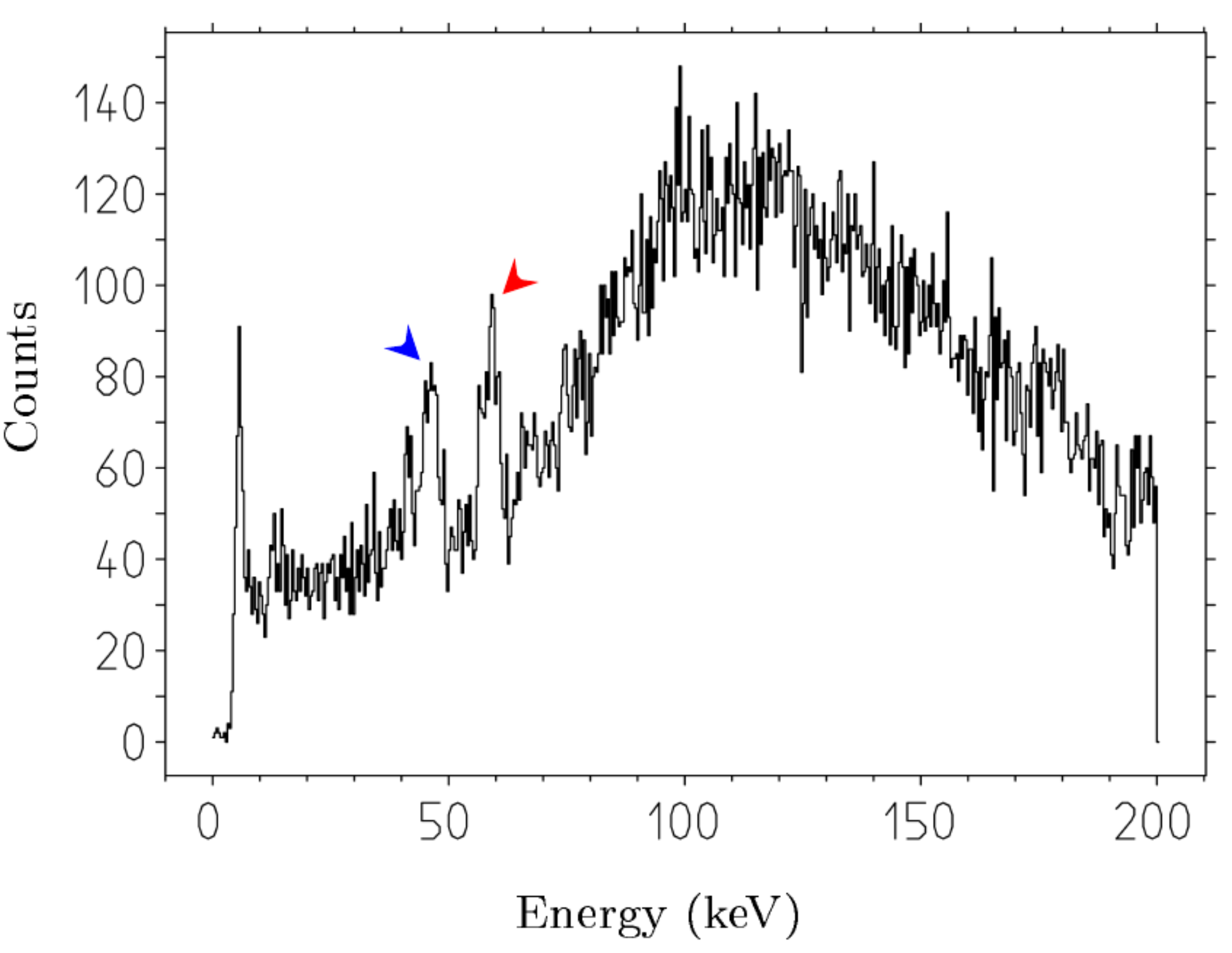}
            }
            \subfigure[Slow pulses (from glued part)]{
              \includegraphics[width=8cm]{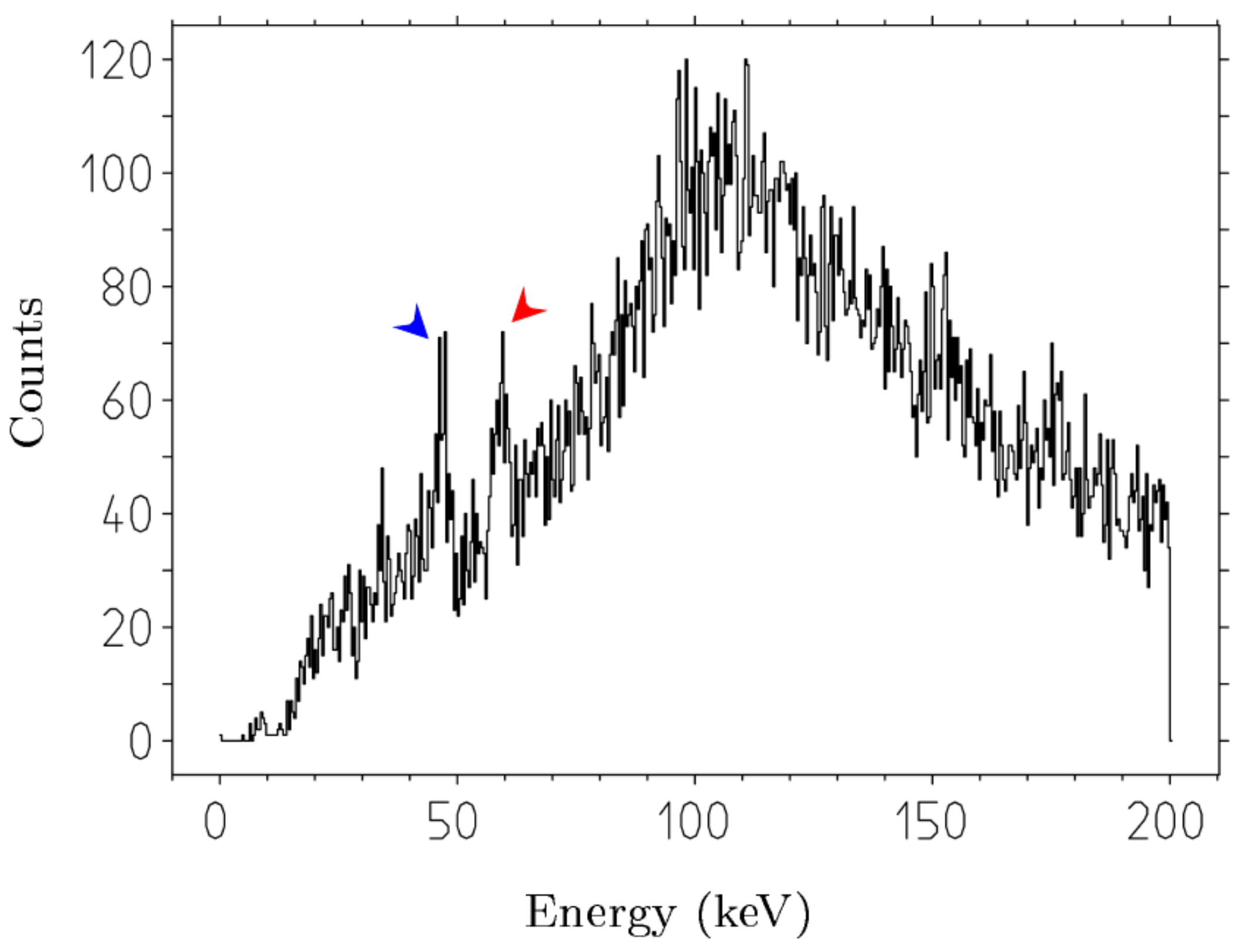}
            }
          \end{center}
          \caption{Comparison of the spectra of pulses originating from the different halves of the crystal. In both crystals, the resolution is $\approx\unit[5]{\%}$ for the $\unit[60]{keV}$ peak.}\label{fig:C14_cut_spectra}
        \end{figure}

        The spectrum which was recorded with the uncut crystal shows two peaks clearly distinguishable from the background, one from the \unit[60]{keV}-source, the other one corresponding to the escape energy from the  tungsten-L-shell at \mbox{\unit[51]{keV}} (see Fig. \ref{fig:C14uncut}).
        Separating the events recorded with the glued detector via their rise times, two spectra from the two absorbers were obtained (see Fig. \ref{fig:C14_cut_spectra}). Both peaks (51 and \unit[60]{keV}) are still separable in the spectra of the components of the glued detector. The resolution even improved from $\approx\unit[7.5]{\%}$ with the entire crystal to $\approx\unit[5]{\%}$ for each of the halves.\par
        In comparison to the uncut crystal, the threshold for measuring pulses could be reduced due to less noise in the measurement electronics: Therefore the spectrum for the thermometer carrier starts at \unit[5]{keV} instead of \unit[10]{keV}. The energy threshold for pulses from the glued-on absorber is increased from \unit[5]{keV} to \unit[15]{keV} in comparison to the thermometer carrier. This is due to a calibration effect: Particles of the same energy create lower pulse heights if being absorbed in the glued-on absorber than if being absorbed in the thermometer carrier. Therefore a certain energy band cannot be measured via the glued-on absorber while measurement with the thermometer carrier is still possible.
      \subsubsection{Calculation of thermal couplings}
        The data of a pulse recorded with the one-crystal-setup was fitted to the model of Equation \eqref{eqn:ModelSingle}. The volume of the crystal was measured and the thermal capacity calculated according to~\cite{tinkham} so that the couplings $G_{b}$ and $A_{f}$ to the heat bath and the thermometer film could be extracted.\par
        The values of these couplings were assumed to remain unaffected by the cutting of the crystal for the second experiment. Using the parameters obtained from the first experiment, it was possible to fit the pulses of the second experiment with Equation \eqref{eqn:ModelAComplex} and to extract the couplings $A_{a}$ and $A_{g}$, which are displayed in Table \ref{tab:transitions}.\par
        The amount of phonons that passed through the glue was lower, but still sufficiently high to measure a spectrum. This decrease in phonon signal quality is tolerable in view of an increase of scintillation light.
        \begin{table}
          \begin{center}
            \begin{tabular}{|r|r|}
              \hline
              $A_{f}$&$\unit[\left(5.5772\pm0.0093\right)\cdot10^{-3}]{m^{3}\;s^{-1}}$\\
              \hline
              $A_{a}$&$\unit[\left(0.82\pm0.34\right)\cdot10^{-4}]{m^{3}\;s^{-1}}$\\
              \hline
              $A_{g}$&$\unit[\left(1.797\pm0.090\right)\cdot10^{-4}]{m^{3}\;s^{-1}}$\\
              \hline
            \end{tabular}
          \end{center}
          \caption{Transition properties of the glue in the proof-of-principle experiment}\label{tab:transitions}
        \end{table}
  \section{Glued detectors for Gran Sasso}
    In the present Run 31 of the CRESST-II experiment, three detectors with thermometer carriers and glued-on absorber crystals consisting of \cawo and a fourth glued detector made out of ZnWO$_{4}$ are included in the Gran Sasso setup.
    \subsection{Carrier crystals for Gran Sasso detectors}
\begin{figure}
        \centering
        \includegraphics[width=7cm]{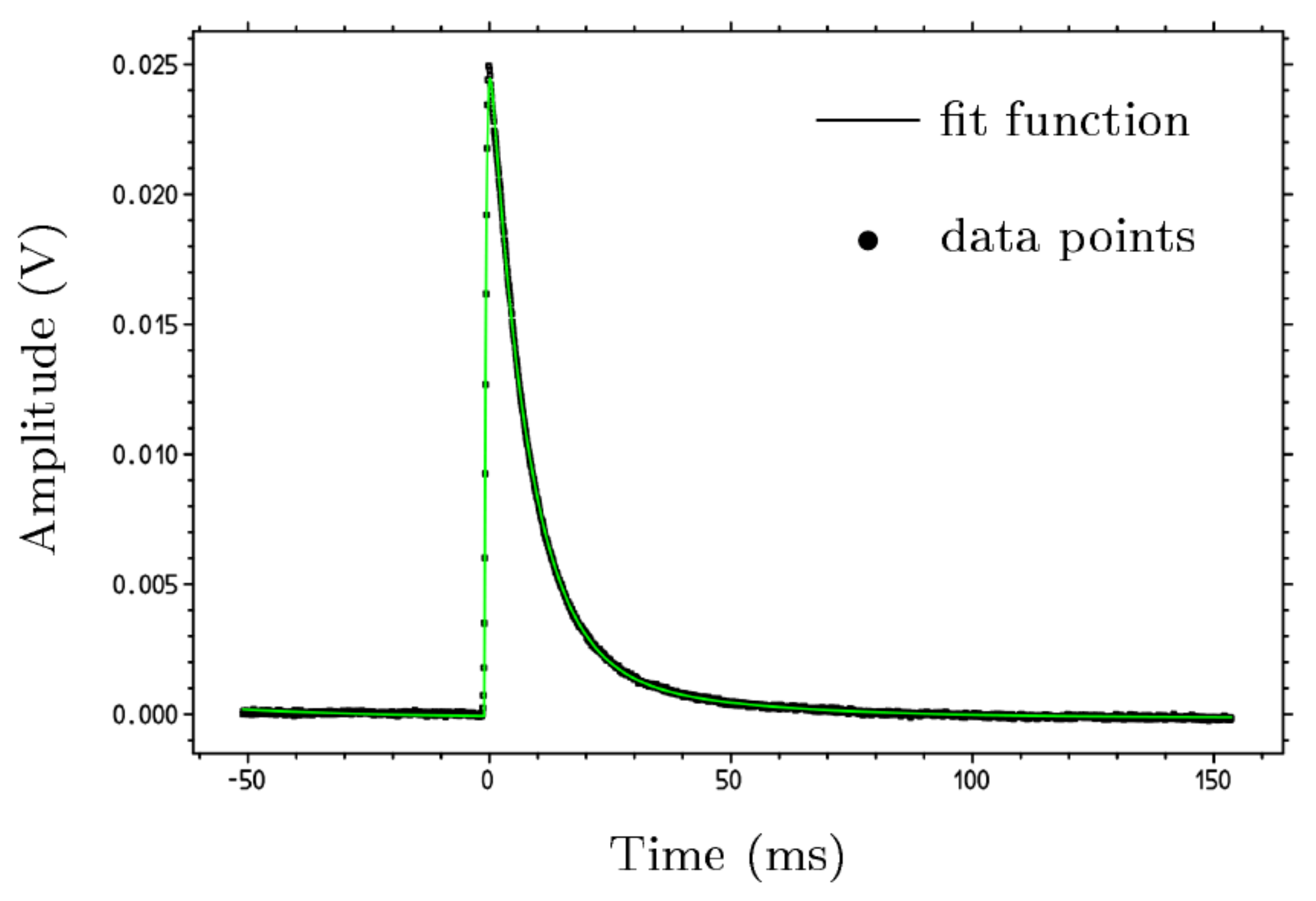}
        \caption{Pulse taken with thermometer carrier WI220. Dots represent data and lines the fitted curves.}\label{fig:WI220} 
      \end{figure}

\begin{figure}
        \centering
        \includegraphics[width=7cm]{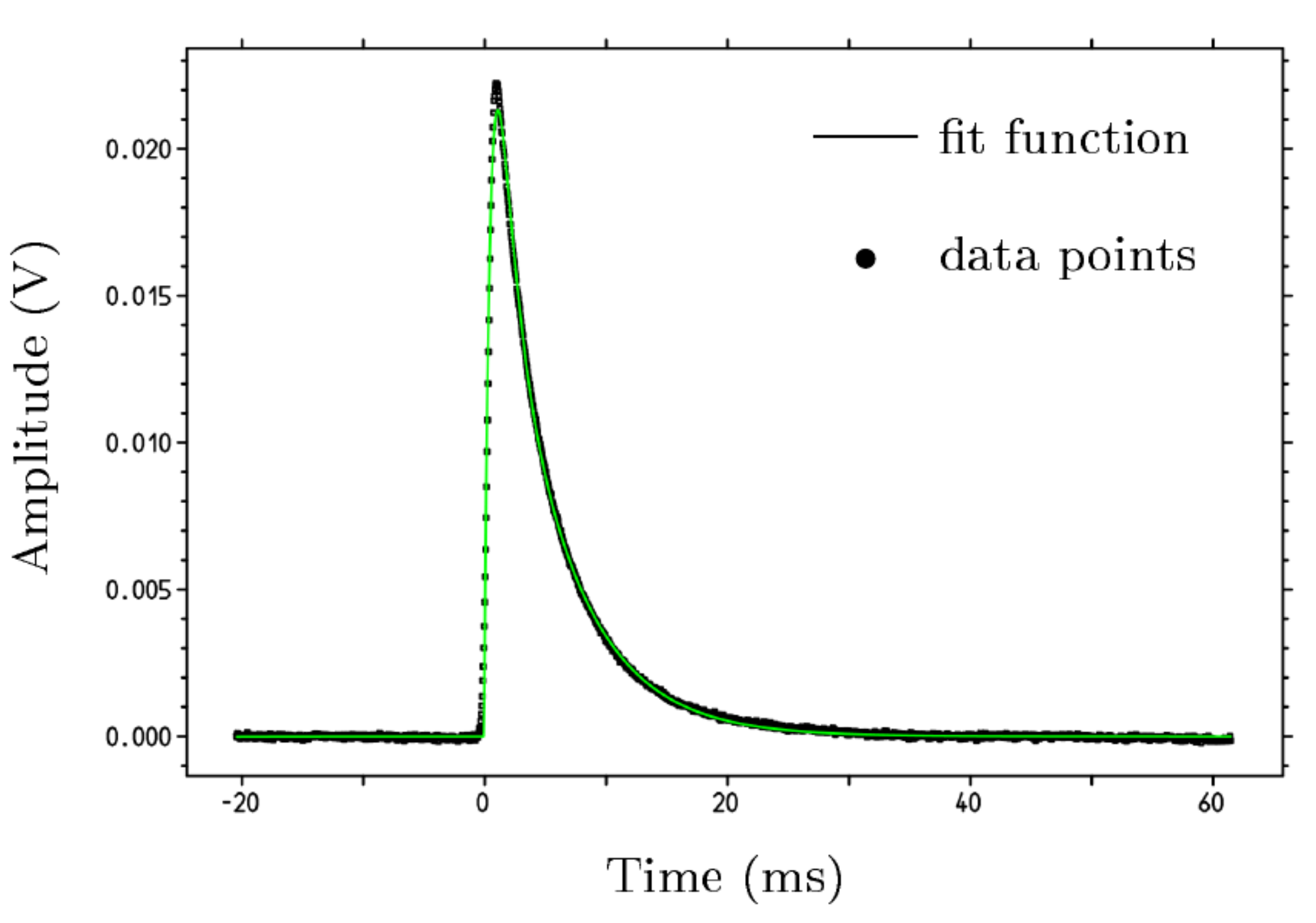}
        \caption{Pulse taken with thermometer carrier WI225. Dots represent data and lines the fitted curves.}\label{fig:WI225} 
      \end{figure}

\begin{figure}
        \centering
        \includegraphics[width=7cm]{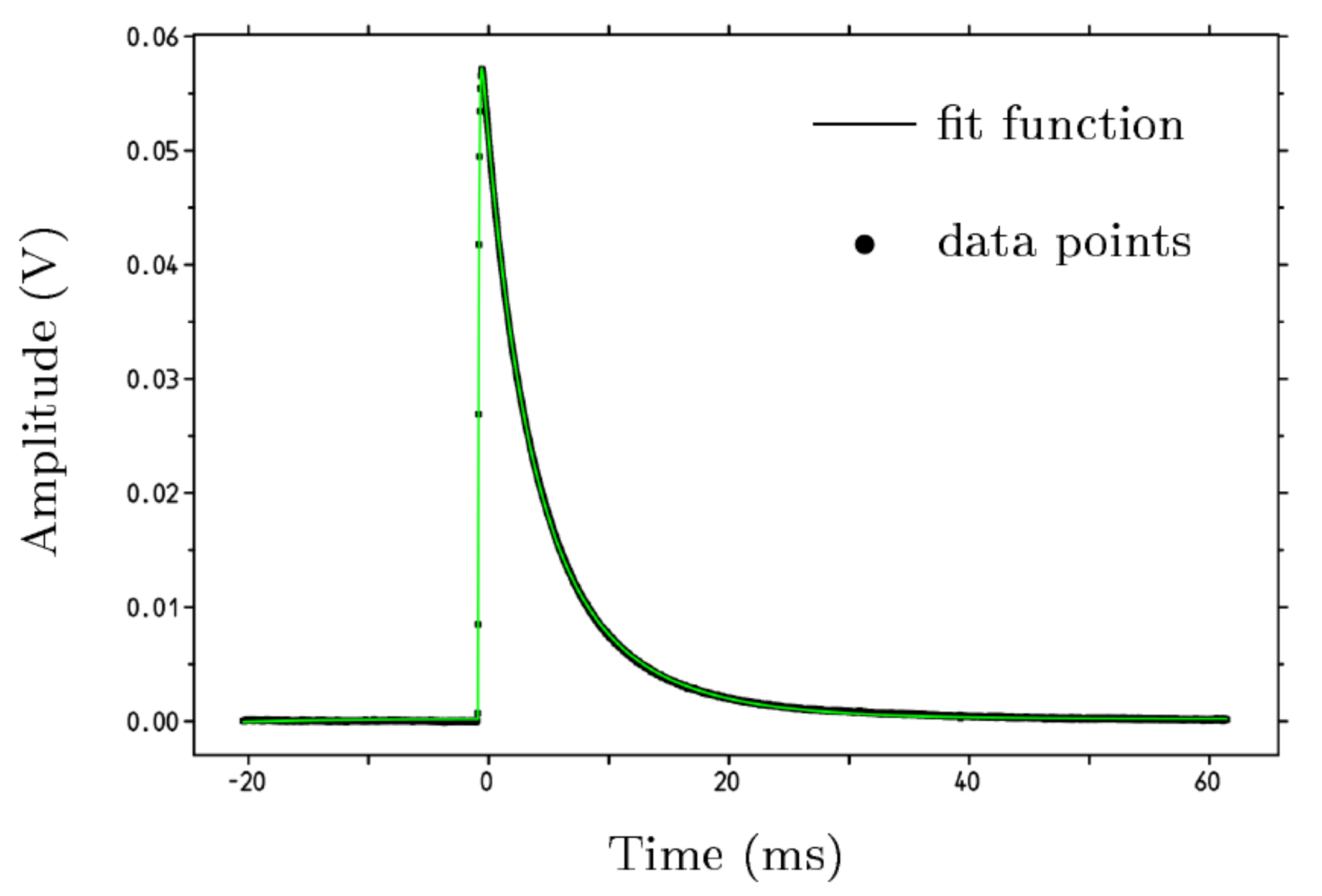}
        \caption{Pulse taken with thermometer carrier WI228. Dots represent data and lines the fitted curves.}\label{fig:WI228} 
      \end{figure}

      In order to determine characteristics of the carrier crystals WI220, WI225 and WI228, each of them was used to take a $^{57}$Co spectrum before being glued. Figures \ref{fig:WI220} to \ref{fig:WI228} show examples of  pulses taken with the carrier crystals.
      Fitting the pulses of the three detectors to the model of Equation \eqref{eqn:ModelSingle} resulted in the characteristics shown in Table \ref{tbl:WI22X}.
      \begin{table}
        \begin{center}
          \begin{tabular}{|r|r@{$\pm$}l|r@{$\pm$}l|r@{$\pm$}l|}
            \hline
             &\multicolumn{2}{|l|}{\textbf{WI220}} & \multicolumn{2}{|l|}{\textbf{WI225}} & \multicolumn{2}{|l|}{\textbf{WI228}}\\
            \hline
            $\mathbf{\tau_{r}}$&$0.3741$&$\unit[0.0011]{ms}$ & $0.7222$&$\unit[0.0012]{ms}$ & $0.0803$&$\unit[0.0011]{ms}$ \\
            $\mathbf{\tau_{n}}$&$7.716$&$\unit[0.016]{ms}$ & $0.8679$&$\unit[0.0018]{ms}$ & $3.1602$&$\unit[0.0044]{ms}$\\
            \hline
          \end{tabular}
        \end{center}
        ~
        \caption{Characteristics (see Equation \eqref{eqn:ModelSingle}) of the pulses measured with the thermometer carrier before gluing.}\label{tbl:WI22X}
      \end{table}
    \subsection{Gluing}
      Following the reference measurement, the thermometer carriers have been glued to cylindrical absorber crystals of \unit[40]{mm} diameter and height. The masses of the crystals are $\approx\unit[310]{g}$ for \cawo and $\approx\unit[400]{g}$ for ZnWO$_{4}$, respectively.\par
      Two different glues have been used: One detector was glued with Araldite 2011, two others with Epo-Tek 301-2\textregistered\footnote{Epo-Tek is a registered trademark of Epoxy Technology, Inc.}. Both glues have been tested at MPI and it has been observed that Araldite 2011 is itself a scintillator under UV light~\cite{karoaraldit}, while Epo-Tek has a lower viscosity. The denser Araldite had to be spread between the two crystals by using a mechanical press while Epo-Tek spreads itself by the weight of the thermometer carrier.\par
      These glued detectors have been installed in the Gran Sasso setup for Run 31 and are currently in use.
  \section{Conclusion}
    Detectors with glued thermometers have proven to operate in a stable manner and to deliver pulses which can be used to measure a spectrum. They bring many advantages in terms of fabrication of the detectors without significantly degrading the energy spectrum.\par
    Nevertheless there is a loss of signal quality in the phonon channel. In the case of CRESST, this loss can be accepted, as the increase in scintillation light is expected to improve the discrimination capabilities.\par
    Gluing thermometers onto scintillator crystals may provide a possibility to use scintillator materials which up to now have not been suitable for experiments like CRESST because of difficulties in depositing a thermometer. This might enlarge the choice of scintillating absorber materials for CRESST and the upcoming EURECA experiment.
  \section{Acknowledgements}
    I would like to thank the CRESST-group of the Technische Universität München for providing a sample of EpoTek glue.
  \bibliographystyle{elsart-num}
  
\end{document}